# Artificial Intelligence for Technical Debt Management in Software Development

Samia Alam Binta
Department of Software Engineering,
LUT University
Lappeenratna Finland
samia.alam@student.lut.fi

Savita Kaushal
Department of Software Engineering,
LUT University
Lappeenratna Finland
savita.kaushal @student.lut.fi

Srinivas Babu Pandi
Department of Software Engineering,
LUT University
Lappeenratna Finland
srinivas.babu.pandi@student.lut.fi

*Abstract*—Technical debt is a well-known challenge in software development, and its negative impact on software quality, maintainability, and performance is widely recognized. In recent years, artificial intelligence (AI) has proven to be a promising approach to assist in managing technical debt. This paper presents a comprehensive literature review of existing research on the use of AI powered tools for technical debt avoidance in software development. In this literature review we analyzed 15 related research papers which covers various AI-powered techniques, such as code analysis and review, automated testing, code refactoring, predictive maintenance, code generation, and code documentation, and explores their effectiveness in addressing technical debt. The review also discusses the benefits and challenges of using AI for technical debt management, provides insights into the current state of research, and highlights gaps and opportunities for future research. The findings of this review suggest that AI has the potential to significantly improve technical debt management in software development, and that existing research provides valuable insights into how AI can be leveraged to address technical debt effectively and efficiently. However, the review also highlights several challenges and limitations of current approaches, such as the need for high-quality data and ethical considerations and underscores the importance of further research to address these issues. The paper provides a comprehensive overview of the current state of research on AI for technical debt avoidance and offers practical guidance for software development teams seeking to leverage AI in their development processes to mitigate technical debt effectively.

## I. Introduction

Technical debt can be defined as the cost software has to pay in the long run due to the compromise or short-cut took by the developers knowingly or un-knowingly to meet the short-term goals [1]. Since technical debts (TD) is viewed in a negative sense, the question arises why TDs even exist in the first place. Technical debts fulfill short-term goals which might become an issue in the long-term sustainability. As software development has become increasingly complex, addressing technical debt has become more challenging, with developers often making trade-offs that result in suboptimal code or design. This has led to an increased interest in leveraging artificial intelligence (AI) to assist in mitigating technical debt. There are so many open-source tools utilizing AI techniques which help in identifying the technical debts and means to mitigate them.

Tools like *SonarQube and CodeClimate* analyses the source code to identify technical debts such as code smells and duplication [2]. Automated testing tools like *Selenium and Appium* identifies technical debts in the testing scripts [3]. Code generation tools like *Yeoman and Spring Initializr*, generates code template which helps in avoiding the technical debts [1]. Documentation tools like *Doxygen and Javadoc* generated the documentation for code automatically which avoids the technical debts that might arise due to lack of documentation [1].

This literature review will cover various AI-powered techniques, their effectiveness in addressing technical debt, and the benefits and challenges of using AI for technical debt avoidance. Previous studies explain a specific kind of technical debt and how a specific tool handles the technical debts. The paper will try to summarize all the previous research done on the topic and various AI tools to provide insights into the current state of research on this topic. In doing so, identifying any gaps and opportunities for future research, and offer practical guidance for software development teams seeking to leverage AI in their development processes to mitigate technical debt effectively.

## II. Methodology

In this section, we will discuss how the research is conducted. The research method we are going to implement for this study is literature review. This literature review is used to collect information about the technical debts, AI techniques, and tools to manage technical debts.

This literature review studies the 15 previous studies which highlight the technical debts and various AI tools. The research in this paper is focused on answering the following research questions:

*Research Questions*

**RQ1**: What is a technical debt and what are the different types of technical debts?
**RQ2**: How AI techniques can be used to identify technical debts
**RQ3**: How technical management tools use AI techniques to manage technical debts and what are the complications that arise due to the use of AI tools.

## III. Classification of Technical Debts

According to Alves, there are so many different types of technical debts [4].

- **Design Debt**: Technical debt incurred due to the bad initial design which might result in poor performance of the system.
- **Code Debt**: Technical debt due to bad source code or developing the code without adhering to coding standards which makes it difficult for further maintenance.
- **Architecture Debt**: Technical debt arises due to poor design choices or a lack of modularity in the overall software architecture which might make it difficult to add new features to the system.
- **Infrastructure Debt**: Technical debt that arises from outdated or inadequate infrastructure or deployment processes which results in issues during deployment.
- **Testing Debt**: Cost incurred on the system as a result of inadequate or insufficient testing which leads to buggy software.
- **Documentation Debt**: Cost incurred on the system during maintenance due to lack of documentation or code comments.
- **Defect Debt**: Technical debt because of willful ignorance of known defects or bugs during software development.
- **Requirement Debt**: Technical debt because of incomplete or unclear requirements before/during the development which results in unusable system.
- **People Debt**: Technical debt because of human errors, like miscommunication between developers or miscommunication between development teams.

The research summarized by Alvis [4] answers our first research question (RQ1). Although there are a few more types of technical debts, we mentioned only relevant types for this study. Relevance is dependent on how these types of technical debts can be identified and managed by AI.

## IV. TECHNICAL DEBT DETECTION USING AI TECHNIQUES

Identifying a technical debt in the system is the first step of TD management. This chapter explains the different AI techniques that are used in detecting technical debts.

### A. Natural Langugae Processing (NLP)

Natural Language Processing (NLP) algorithms analyzes various sources of texts like code comments, commit messages, and issue trackers to identify the technical debts. As most of the IT organizations tend to use agile methodology for software development, research has been conducted by Quintana [5] on using Machine learning technique to identify documentation debt. One of the advantages of agile methodology is the flexibility it offers to the developers as it allows requirements to change over time and developers can change the design instantly. According to Quintana, developers fail to keep up with the documentation with quick changes in the requirements causing technical debt. By leveraging AI technique *Natural Language Processing* (**NLP**), documentation debt can be identified. NLP can be broadly divided into 3 categories which are lexical analysis, semantic analysis, and discourse analysis. The research conducted by Quintana utilizes the potential of NLP to design a text processing tool to identify the technical debt in the documentation and in turn reduce the impact of technical debt.

### B. Static Analysis

Static Analysis is another AI technique which can be used to detect technical debts, more specifically code debts. Static analysis is a method for analyzing software code and spotting potential problems without running it. Static analysis examines the code for potential flaws, best practices violations, and other problems that could result in technical debt and mark them as high or low TD based on its impact on the overall system. Research conducted by Tsoukalas [6] analyzes various algorithms which use static analysis to identify the technical debts in software systems. The research concludes a superior classifier algorithm is best in static code analysis in identifying technical debts.

Research paper submitted by Rantala [7], uses both NLP and static analysis in identifying TDs. According to Rantala, combining both NLP and static analysis improves the accuracy of identifying the technical debt, other than using them separately. In this study NLP is used to understand the developer's messages inside the code and static analysis to analyze the code which identifies the code debts accurately. Static analysis checks for repetitive blocks of code or any misalignment from the standard coding practices. NLP then analyzes the commit messages or code comments to understand the reason for violation (if any) and identifies the technical debt as high or low.

### C. Deep Learning

Deep learning is another AI technique to identify technical debts. An algorithm analyzes large volumes of data and understands the pattern (which indicates technical debts) in the dataset. In a study to identify the self-admitted technical debts (SATD), deep learning algorithms are used by Zampetti [8]. The researchers combined two of the most popular deep learning algorithm Convolutional neural networks (CNN) and Recurrent neural networks (RNN) to identify and remove the SATDs from the online code base. CNN is used to analyze the commit messages and comment, whereas RNN is used in analyzing the code patterns and structure of the code. This study not only identifies the technical debts like code smells, but it is also 55% successful in removing the SATDs.

In another study by Aversano [9], deep learning algorithm is used to extract the technical debts from source code repositories. Technical debt is quantified by measuring various software quality metrics. This study not only identifies the current technical debts, but it also estimates the future technical debts which may degrade the system. This study is mainly focused on addressing design debts which are more concerning for the software maintenance. There are some other algorithms in deep learning like feature extraction, predictive modelling, anomaly detection etc., can identify technical debts in large datasets.

### D. Cognitive Biasing

Cognitive biasing is a kind of counter technique of technical debt. According to the research conducted by Borowa [10], cognitive biasing introduces architectural debts in the system. By analyzing the cognitive biases, we can reduce or completely avoid architectural debts. He also proposed debiasing techniques to tackle the architectural debts due to cognitive biasing.

Although, there are so many AI techniques being used to identify technical debts in the system. I've listed broad categories of methods, which I identified from reviewing the 6 research papers [5,6,7,8,9,10]. This answers our 2nd research question.

## V. TECHNICAL DEBT MANAGEMENT USING AI TOOLS

Technical debt management (TDM) starts with identifying the technical debt, then prioritizing, tracking, and managing the debts. The techniques mentioned in chapter iv are useful in identifying the techniques and at some way across the TDM, it requires some tools to manage the technical debts automatically. By leveraging machine learning algorithms, Certain tools like sonarqube, CAST, etc, are developed to manage different types of technical debts. This chapter discusses how different types of technical debts are managed automatically.

### A. Managing *code* and *design* debts

There are many tools like SonarQube, CAST, Sonargraph, codeinspector, etc., available to manage code/design debts. A new tool was developed by Diogo Pina [11] and others who participated in the research, naming it sonarlizer Xplorer. This tool combines GitHub Xplorer and Sonarqube to manage code debts. GitHub Xplorer mines the public repositories and sonarqube analyzes the code and identifies the technical debts measures the code metrics from the public repositories. During the tenure of 4 months doing research, this team analyzed over 46,000 Public java projects to find the technical debts and measure the code metrics.

Sonarqube is a code analysis tool, goes through the code, commit messages and code comments to identify and measure the technical debts. Then it classifies the technical debt based on their severity. Of the 46,000 java projects it analyzed in the research, it has some interesting findings can be found in figure 1.

| Severity | # TD items | ncloc | # Projects |
|---|---|---|---|
| Blocker | 467k | <1k | 27k |
| Critical | 2.4M | 1k - 10k | 14k |
| Major | 5.2M | 10k - 100k | 4.6k |
| Minor | 6.6M | 100k - 500k | 390 |
| Info | 400k | >500k | 17 |

Fig. 1. *Results of the research based on 46,000 projects [11]*

Sonarqube usually finds the technical debts from the code comment by developers which are considered to be self-admitted technical debts and it also finds technical debts after running the code. These kinds of debts are mentioned as ncloc in Fig. 1. Along with sonarqube, there are similar tools to manage code debts such as, CAST, N Depend, Squore, Code Inspector, and SymfonyInsight. All these tools are useful in managing code debts.

*Table 1: Different AI technical debt management tools*

| Tool | Working Methodology | Benefits | Drawbacks |
|---|---|---|---|
| SonarQube | Static code analysis to identify and manage code debts | Open-source, supports multiple codng languages | Inaccuracies in identifying technical debts or prioritizing with open-source version |
| CAST | Uses machine learning algorithms to analyze code and measure code quality metrics | More comprehensive code analysis, can be integrated with most used IDEs | False positives, Not an open-source tool |
| CodeInspector | Static code analysis to identify technical debts and dashboard to track the debts | Offers comprehensive analysis of code quality and technical debt, Provides detailed reports and code metrics | Relatively new tool and not many features in he dashboard |
| N Depend | Static code analysis to identify technical debts and dashboard to track the debts | Offers comprehensive analysis and visualization of code quality and technical debt, Supports many programming languages and frameworks | Complex set-up and not so user-friendly as compared to the other tools |
| Squore | Comprehensice code analysis | Offers comprehensive analysis for multiple programming languages and visualization of code quality and technical debt. | Expensive when compared to other related tools |
| Symfony Insight | Static code analysis | Identifies TDs more comprhensively on PHP frameworks | Supports only PHP language which manages design debts |

Table 1 explains the advantages and disadvantages of the most popular code and design technical debt management tools.

### B. Managing *Architecture* and *infrastructure* debts

Architecture and infrastructure debt is not directly related to the code debt, but rather related to architectural decisions which might create issues in the later stages of the software development life cycle.

Based on the research conducted by Lenarduzzi, changing the architectural style from traditional monolithic system to microservice architecture will reduce the technical debt [12]. So, choosing right architecture in the project start-up solves most of the technical debs which may arise in the future. Another research by the Lenarduzzi gives the comprehensive analysis of the available technical debt management tools [13]. ARCAN is one such tool which specializes in managing architectural debts. ARCAN (Architecture Analysis) use both static and dynamic code analysis tool to analyze the code to detect code and architectural issues and inform developers the potential measure to mitigate the architectural debts.

iPlasma is another open-source tool to manage technical architecture debts. It uses static code analysis to measure the modularity and coupling of software modules. It compares it with standard code metrics and informs the developers about potential technical debts and creates a visualization dashboard to track the technical debts. Zhang [14] employed iPlasma to create a new tool named Delesmell to handle technical debts. This research employed various tools including iPlasma and machine learning algorithms to design delesmell. By combining multiple techniques, the researchers designed a single technical debt management tool. It used iPlasma to find the architectural issues within the system and convolutional neural network (CNN) branch, gate recurrent unit (GRU)-attention branch for classifying the debts. The final classification is conducted by supporting vector machine (SVM). Finally, Delesmell uses code refactoring tool to manage the technical debts.

*Table 2: Differences between ARCAN and iPlasma*

| Parameter | ARCAN | iPlasma |
|---|---|---|
| Methodology | ARCAN uses graph theory and network analysis techniques to identify architectural technical debts and potential design flaws in the system. | iPlasma uses a set of code metrics and visualization techniques to identify potential design flaws and technical debt in software systems |
| Advantages | Can analyze large and complex software systems, Identifies architectural debts and potential design flaws. | Identifies modularity and complexity issues, Provides insights into software quality and technical debt, Offers refactoring suggestions based on code analysis |
| Disadvantages | Time consuming for a large system, High chance of getting a false positive | Limited for analyzing only small systems. |

Table 2 highlights the differences between ARCAN and iPlasma, two different tools for managing architectural debts. According to Lenarduzzi [13] other tools that handle architectural debts are Coverity, Findbugs, FxCopAnalyzer, Jsprit, Scitool, Codescene. The research paper gives a comprehensive analysis of these tools.

C. Managing **Testing** debts

Testing debt is the cost incurred on the system due to incomplete or inaccurate software testing. Testing debt can be mitigated or managed easily by elaborative testing and finding all the relevant test cases. Artificial intelligence has so much potential in this area which can make any software a testing debt-free. Authors of research paper [15] experimented with the popular automated testing tools in the industry and drew a comparison between them.

Selenium is one such automatic framework of manual testing for functional and regression testing, and it supports a wide range of web browsers and programming languages. It provides a set of tools and libraries that allow developers to automate web-based user interface testing, including simulating user interactions, capturing screenshots, and generating test reports.

But in order to manage technical debt selenium can be integrated with AI-based tools generate the test cases automatically. A few of those tools are *Appvance IQ, Functionize*, and *Mabl*.

- **Functionize** generates test cases and executes selenium test scripts automatically.
- **Appvance IQ** analyzes selenium test results and identify patterns in the test failures which can provide insights into common failure scenarios and help prioritizing test efforts.
- **Mabl** also analyzes the selenium test results and identify patterns in the test failures which helps in identifying the root cause for failures and suggest areas for improvements in further testing process.

**Testim** is another AI-based automatic testing tool which can generate testcases automatically. It uses machine learning algorithms to learn from user behavior and automatically generate test cases. It eliminates the need of having a separate tester to prepare test cases. It also provides a visual editor that allows users to create and modify tests without writing code, and integrates with popular tools such as Jira, Jenkins, and Slack.

*Table 3: Differences between different AI-based testing tools with the potentiality to avoid testing debts.*

| Tool | Working Methodology | Benefits | Drawbacks |
|---|---|---|---|
| Selinium | Run test cases automatically usign testing scripts | Open Source, can be customized according to the system's need. | Complex set-up and requires technical expertise in utilizing all the features. |
| Testim | Completely automatic testing platform using AI techniques | Fully automatic, easy to use, doesn't requrie technical expertise. | Not suitable for bigger systems, not open-source |

According to Lal [15] the other tools that can be used to manage testing debts are Testim, Nibbler, Pingdom, and wave accessibility tools. Table 3 summarizes the key differences between selenium and Testim testing tools about the working methodology, Advantages, and disadvantages of employing them for mitigating testing debts.

D. Managing **Documentation** debts

Documentation debt refers to the cost incurred on the system because of inadequate or incomplete documentation over time, resulting in a backlog of documentation work that needs to be done. Going through the documentation is mandatory to understand how the system works and it eases the problem of understanding someone else's code. Documentation is a manual task if we think of it, but there are individual areas where we can automate micro tasks which can help in managing the document debt. For example, automated comments for coding. Although there is no solid research conducted on mitigating the documentation debt, I found few tools which help in avoiding the documentation debts, but most of them are platform specific.

**JavaDoc** is tool for automatic comment generation for java source code. Although it's not a direct tool for managing technical debt, it helps in mitigating the document debt by automatic generation of comments in the source code.

**Sphinx** is a document generator for generating technical documentation for Python, C+ and Javascript source codes.

**Sandcastle** is similar to JavaDoc but for .Net code. It analyses XML comments in the code to generate technical documentation.

Roxygen JavaDoc's equivalent to R programming to avoid document debt.

According to my research **Doxygen** is the only AI-based tool, that can generate documentation for multiple programming languages. It primarily supports C++, but it also works for all the other source code.

## VI. CONCLUSION

In this Paper, we reviewed 15 different research articles and other open-source technologies to get a comprehensive understanding of different types of technical debts. We also reviewed a few research articles utilizing artificial intelligence techniques to manage technical debts. Chapter III highlights the classification of technical debts and defines each type of debts. Chapter IV discusses the various machine learning techniques to identify the technical debts in the software. Based on literature review we identified Natural Language Processing, Static analysis, Deep learning, and Cognitive biasing are the popular AI techniques used to identify technical debts in software which is the first step in technical debt management. Chapter VI lists out the popular tools that are in use for technical debt management. This chapter also details how different types of technical debts are handled using AI techniques and tools.

### FUTURE SCOPE

Since **A**rtificial **I**ntelligence usage in the software engineering project is growing at an exponential rate, there is so much scope for AI algorithms in technical debt Management. Based on our survey, there is less research available in document debt management.